\documentclass{jfm}
\usepackage{graphicx}
\usepackage{epstopdf, epsfig}
\usepackage{hyperref}

\shorttitle{Deformations of a pre-stretched and lubricated finite elastic sheet}
\shortauthor{E. Boyko, R. Eshel, K. Gommed, A. D. Gat and M. Bercovici}

\title{Elastohydrodynamics of a pre-stretched finite elastic sheet lubricated by a thin viscous film with application to microfluidic soft actuators}

\author{Evgeniy Boyko, Ran Eshel, Khaled Gommed,
  Amir D. Gat \corresp{\email{amirgat@technion.ac.il}}
 \and Moran Bercovici \corresp{\email{mberco@technion.ac.il}}}

\affiliation{Faculty of Mechanical Engineering, Technion -- Israel Institute of Technology, Haifa, Israel}
\begin{document}
\maketitle
\begin{abstract}
The interaction of a thin viscous film with an elastic sheet results in coupling of pressure and deformation, which can be utilized as an actuation mechanism for surface deformations in a wide range of applications, including microfluidics, optics, and soft robotics. Implementation of such configurations inherently takes place over finite domains and often requires some pre-stretching of the sheet. Under the assumptions of strong pre-stretching and small deformations of the lubricated elastic sheet, we use the linearized Reynolds and F$\mathrm{\ddot{o}}$ppl\textendash von K$\mathrm{\acute{a}rm\acute{a}n}$ equations to derive closed-form analytical solutions describing the deformation in a finite domain due to external forces, accounting for both bending and tension effects. We provide a closed-form solution for the case of a square-shaped actuation region and present the effect of pre-stretching on the dynamics of the deformation. We further present the dependence of the deformation magnitude and timescale on the spatial wavenumber, as well as the transition between stretching- and bending-dominant regimes. We also demonstrate the effect of spatial discretization of the forcing (representing practical actuation elements) on the achievable resolution of the deformation. Extending the problem to an axisymmetric domain, we investigate the effects arising from nonlinearity of the Reynolds and F$\mathrm{\ddot{o}}$ppl\textendash von K$\mathrm{\acute{a}rm\acute{a}n}$ equations and present the deformation behavior as it becomes comparable to the initial film thickness and dependent on the induced tension. These results set the theoretical foundation for implementation of microfluidic soft actuators based on elastohydrodynanmics. 
\end{abstract}

\section{Introduction}
Elastohydrodynamic interaction between an elastic substrate and a thin liquid film is of interest in the study of various natural processes, such as passage of air flow in the lungs \citep{grotberg2004biofluid} and geological formation of laccoliths \citep{michaut2011dynamics}, as well as in the dynamic control of elastic structures for applications in soft robotics, adaptive optics and reconfigurable microfluidics \citep{thorsen2002microfluidic,chronis2003tunable,kim2013soft}.
In particular, the case of a viscous fluid confined between an elastic sheet and a rigid surface has been extensively studied in the context of viscous peeling \citep{hosoi2004peeling,lister2013viscous,hewitt2015elastic}, suppression of viscous fingering instabilities \citep{pihler2012suppression,pihler2013modelling,al2013two,pihler2014interaction},  impact mitigation \citep{tulchinsky2016}, elastohydrodynamic wakes \citep{wake}, and dynamics of wrinkling of a lubricated elastic sheet \citep{kodio2016lubricated}.

\begin{figure}
 \centerline{\includegraphics[scale=1]{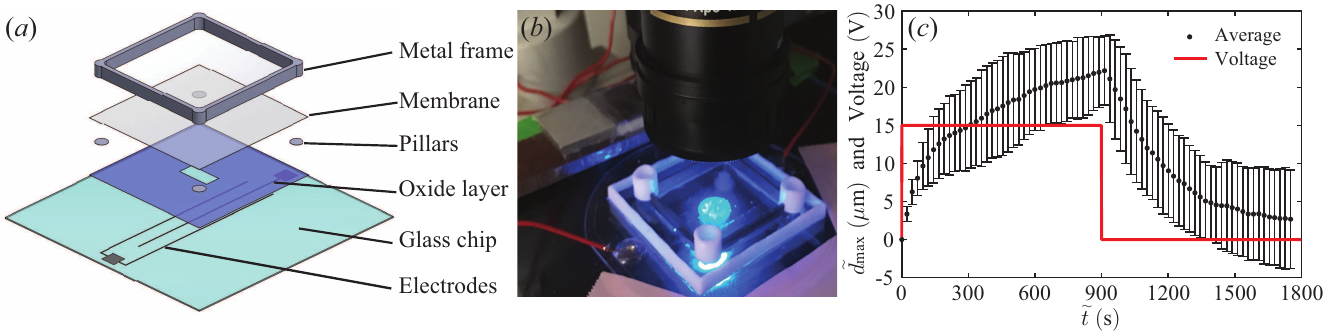}}
\caption{(a) Illustration and (b) image of an experimental setup for deformation of an elastic sheet, actuated by non-uniform electro-osmotic flow. The configuration consists of a 12 $\mu$m thick low-density polyethylene sheet stretched over a rigid frame that is supported by 100 $\mu$m tall pillars from a rigid glass substrate, forming a Hele-Shaw chamber. The glass surface is patterned with three parallel electrodes and coated with an insulating oxide layer, excluding a rectangular actuation region at the center. The chamber is filled with an aqueous solution, and the voltage is set to drive electro-osmotic flow from the outer electrodes to the inner ones, resulting in internal pressure gradients. (c) Experimental results showing the growth and decay of deformation at the center of the elastic membrane, in response to sudden actuation and cessation of the electric field. Error bars indicate a $95\,\%$ confidence on the mean based on six repeats, and the red curve represents the corresponding applied voltage as a function of time.}
\label{F1}
\end{figure}

In the field of microfluidics, where channels are often fabricated from soft materials (e.g. PDMS), there is growing interest in exploring the effect of elasticity on the resulting flow and pressure fields, and several theoretical works have addressed this subject \citep{gervais2006flow,dendukuri2007stop,hardy2009deformation,panda2009temporal,mukherjee2013relaxation,christov2017flow}. For example, \citet{gervais2006flow} used a one-dimensional model, based on the assumption of a linear relation between fluidic pressure and channel deformation, to estimate the effect of elasticity on the pressure field in a shallow elastic micro-channel, showing good agreement with their experimental results. Recently, \citet{christov2017flow} applied the lubrication approximation and the Kirchhoff-Love bending model to derive the relation between flow rate and pressure drop for a deformable shallow micro-channel. In tandem with the effect of elasticity on the flow field, the use of fluidic forces as an actuation mechanism for deformation of elastic substrates has also been exploited in a variety of applications, primarily as an on-chip valving method \citep{unger2000monolithic,grover2006development}, and recently also for other applications including adaptive optics \citep{jeong2004tunable} and soft robotics \citep{shepherd2011multigait}.   

Of particular interest are soft planar microfluidic configurations, which may serve as a platform for re-configurable microstructures. For such configurations, thin elastic sheets (e.g. a $\sim$10 $\mu$m polymer sheet) are a natural choice for maximizing deformations. While tension reduces elastic deformations, it is difficult to implement robust setups without introducing some pre-stretching of the sheets. Thus, pure bending models (e.g. such as the one considered by \citet{rubin2017} in previous work from our group) are insufficient for describing the behavior of realistic systems. Furthermore, accurate prediction of the deformation field requires accounting for the influence of the finite boundaries of the sheet. As an example, and to relate the theoretical model studied in this work to realistic configurations, figures \ref{F1}(a--b) present an experimental setup in which
a thin elastic membrane, pre-stretched on a rigid frame, serves as
the ceiling of a Hele-Shaw chamber. The elastic sheet is actuated
by an internal pressure gradient formed by opposing electro-osmotic
flows (EOF) within the chamber. Additional details
are provided in appendix A of the supplementary material. Figure \ref{F1}(c)
presents experimental results showing that indeed EOF-based deformations
are feasible in practice. We note that since the presented
measurements are focused on a single point, the observed over-damped
response can be easily fitted with various sets of realistic physical parameters and thus cannot be used for proper validation of the model. Such validation
would require an imaging system which would allow much faster data
acquisition as well as imaging of complete surfaces rather than an
individual point. We are currently developing this infrastructure,
which will be reported in the future. 

In this work, we aim
to set the theoretical framework for addressing such configurations,
where finite boundaries, pre-stretching, and fluidic actuation dominate
the physical response of the system. In $\mathsection$ 2, we present the problem formulation and the equations
governing the deformation dynamics, accounting for both bending and stretching, as well as for forces applied either through the non-uniform slip velocity in the fluid or due to the pressure applied directly to the elastic sheet. We provide their scaling and summarize the key assumptions used in the derivation of the model.
Focusing on electro-osmotic actuation, in $\mathsection$ 3 we present
a closed-form solution for the practical case of a square-shaped actuation
region within a finite rectangular domain. We present the effect of pre-stretching
on the steady-state deformation pattern and magnitude, and examine
the timescales for development of pressure and deformation. Using
a Fourier decomposition, in $\mathsection$ 4 we further study the
tradeoff between the amplitude and timescale of deformations and the
attainable spatial resolution, and show their different scaling in tension-
versus bending-dominant regimes. In $\mathsection$ 5, we examine the
effect of spatial discretization of the forcing (representing, e.g.,
actuation electrodes) on the resulting deformation, minimizing the error between the resulting and desired deformation using a least-squares method.
Considering an axisymmetric domain, in $\mathsection$ 6 we employ asymptotic and numerical methods to explore the effects arising from nonlinearity of the Reynolds and F$\mathrm{\ddot{o}}$ppl\textendash von K$\mathrm{\acute{a}rm\acute{a}n}$ equations, and further examine the influence of these effects on the deformation and tension field. We conclude with a discussion of the results in $\mathsection$ 7.

\section{Problem formulation and governing equations}
\begin{figure}
 \centerline{\includegraphics[scale=1]{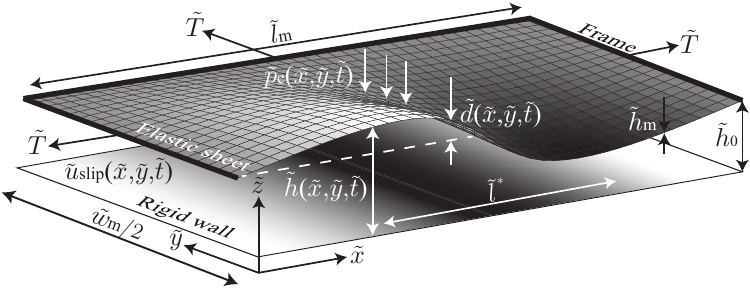}}
\caption{Schematic illustration showing the centerline cross-section of the examined configuration. A thin viscous fluid layer of initial
thickness $\tilde{h}_{0}$ is confined between a rigid surface
and a pre-stretched elastic sheet supported at its boundaries. Non-uniform and time-varying slip velocity $\tilde{\boldsymbol{u}}_{\mathrm{slip}}(\tilde{x},\tilde{y},\tilde{t})$
and external pressure $\tilde{p}_{\mathrm{e}}(\tilde{x},\tilde{y},\tilde{t})$ drive the viscous\textendash elastic interaction and create the deformation field $\tilde{d}(\tilde{x},\tilde{y},\tilde{t})$.}
\label{F2}
\end{figure}
We study the viscous\textendash elastic dynamics of a viscous fluid of density
$\tilde{\rho}$ and viscosity $\tilde{\mu}$ confined between a flat rigid
surface and a pre-stretched elastic sheet of length $\tilde{l}_{\mathrm{m}}$,
width $\tilde{w}_{\mathrm{m}}$, thickness $\tilde{h}_{\mathrm{m}}$, Young's modulus
$\tilde{E}_{\mathrm{Y}}$, and Poisson's ratio $\nu$, as shown in figure \ref{F2}.
We denote dimensional variables by tildes, normalized variables without
tildes and characteristic values by an asterisk superscript. 
We employ a Cartesian coordinate system $(\tilde{x},\tilde{y},\tilde{z})$, and adopt the $\Vert$
and $\bot$ subscripts to denote parallel and perpendicular directions
to the $\tilde{x}-\tilde{y}$ plane, respectively. 

The fluid velocity is $\mathbf{\tilde{\mathrm{\mathit{\boldsymbol{u}}}}}=\left(\boldsymbol{\tilde{u}}_{\Vert},\tilde{u}_{\bot}\right)=\left(\tilde{u},\tilde{v},\tilde{w}\right)$
and fluid pressure is $\tilde{p}$. The total gap between the plates
is $\tilde{h}(\tilde{x},\tilde{y},\tilde{t})=\tilde{h}_{0}+\tilde{d}(\tilde{x},\tilde{y},\tilde{t})$,
where $\tilde{t}$ is time and $\tilde{h}_{0}$ is the initial gap. 
The deformation field $\tilde{d}(\tilde{x},\tilde{y},\tilde{t})$
is induced either due to an external pressure $\tilde{p}_{\mathrm{e}}(\tilde{x},\tilde{y},\tilde{t})$,
acting directly on the elastic sheet, or due to an internal pressure formed by a non-uniform slip velocity
$\tilde{\boldsymbol{u}}_{\mathrm{slip}}(\tilde{x},\tilde{y},\tilde{t})$ on
the rigid surface, which varies over
a characteristic length scale $\tilde{l}^{*}$ in the $\tilde{x}-\tilde{y}$
plane. The characteristic velocities in the $\tilde{x}-\tilde{y}$ plane and $\boldsymbol{\hat{z}}$
direction are respectively $\tilde{u}^{*}$ and  $\tilde{w}^{*}$, and the characteristic pressure, deformation and time are respectively denoted as
$\tilde{p}^{*}$, $\tilde{d}^{*}$, and $\tilde{t}^{*}$.

The most general description for the dynamics of a thin elastic sheet is given by the nonlinear F$\mathrm{\ddot{o}}$ppl\textendash von K$\mathrm{\acute{a}rm\acute{a}n}$ equations \citep{timoshenkoPlates,howell2009applied}, which account for bending and tension forces, external traction, as well as the solid's inertia.   
In this work, we assume that the solid's inertia is negligible and focus on the case of a strongly pre-stretched elastic sheet with isotropic tension $\tilde{T}$, assumed to be much larger than any internal tension, $\tilde{T}_{\mathrm{in}}$, forming in the system during actuation of the sheet. From scaling of the F$\mathrm{\ddot{o}}$ppl\textendash von K$\mathrm{\acute{a}rm\acute{a}n}$ equation that couples the spatial variations in internal tension to Gaussian curvature \citep[p. 176]{howell2009applied}, we find $\tilde{T}_{\mathrm{in}}\sim(\tilde{d}^{*}/\tilde{l}^{*})^{2}\tilde{E}_{\mathrm{Y}}\tilde{h}_{\mathrm{m}}$, and define
\begin{equation}
\alpha=\frac{\tilde{T}_{\mathrm{in}}}{\tilde{T}}=\left(\frac{\tilde{d}^{*}}{\tilde{l}^{*}}\right)^{2}\frac{\tilde{E}_{\mathrm{Y}}\tilde{h}_{\mathrm{m}}}{\tilde{T}}\ll1.\label{alfa}
\end{equation}
Expanding the deformation field and the tension in powers of $\alpha$ and considering the leading order, the nonlinear F$\mathrm{\ddot{o}}$ppl\textendash von K$\mathrm{\acute{a}rm\acute{a}n}$ equations reduce to a single linear plate equation containing bending and pre-stretching terms \citep[see][]{timoshenkoPlates,howell2009applied}
\begin{equation}
\tilde{p}=\tilde{B}\tilde{\nabla}_{\Vert}^{4}\tilde{d}-\tilde{T}\tilde{\nabla}_{\Vert}^{2}\tilde{d}+\tilde{p}_{\mathrm{e}},\label{Elastic balance}
\end{equation}
where $\mathbf{\mathbf{\boldsymbol{\tilde{\nabla}}}_{\Vert}}=\left(\partial/\partial\tilde{x},\partial/\partial\tilde{y}\right)$
is the two-dimensional gradient and $\tilde{B}=\tilde{E}_{\mathrm{Y}}\tilde{h}_{\mathrm{m}}^{3}/12(1-\nu^{2})$
is the bending stiffness. 

We restrict our analysis to shallow geometries, to negligible fluidic inertia,
represented by small Womersley and reduced Reynolds numbers, and to  small elastic deformations,
\refstepcounter{equation}
$$
  \epsilon=\frac{\tilde{h}_{0}}{\tilde{l}^{*}}\ll1,\quad Wo=\frac{\tilde{\rho}\tilde{h}_{0}^{2}}{\tilde{\mu}\tilde{t}^{*}}\ll1,\quad\epsilon Re=\epsilon\frac{\tilde{\rho}\tilde{u}^{*}\tilde{h}_{0}}{\tilde{\mu}}\ll1,\quad\beta=\frac{\tilde{d}^{*}}{\tilde{h}_{0}}\ll1.
  \eqno{(\theequation{a-d})}\label{Shallow}
$$
The first three assumptions (\ref{Shallow}\textit{a--c}) allow us to apply the lubrication
approximation and to obtain a nonlinear Reynolds equation \citep{Leal}, and 
the last assumption, $\beta\ll1$, enables its linearization \citep[see e.g.][]{tulchinsky2016,kodio2016lubricated}.

In $\mathsection$ 6 we relax the assumptions (\ref{alfa})
and (\ref{Shallow}\textit{d}), and explore the effects of nonlinearity of the Reynolds and F$\mathrm{\ddot{o}}$ppl\textendash von K$\mathrm{\acute{a}rm\acute{a}n}$ equations on the deformation and tension fields, considering an axisymmetric geometry.

Based on (\ref{Shallow}), the fluid motion is governed by the lubrication equations \citep{Leal}
\begin{equation}
\tilde{\boldsymbol{\nabla}}\cdot\tilde{\boldsymbol{u}}=0,\quad\mathbf{\tilde{\boldsymbol{\nabla}}_{\Vert}}\tilde{p}=\tilde{\mu}\frac{\partial^{2}\tilde{\boldsymbol{u}}_{\Vert}}{\partial\tilde{z}^{2}},\quad\frac{\partial\tilde{p}}{\partial\tilde{z}}=0.\label{Continuity+Momentum}
\end{equation}
We assume that the fluid is subject to the slip and the no-penetration boundary conditions on the bottom surface. 
Horizontal motion of the elastic sheet is negligible in the small-deformation limit, implying the no-slip and the no-penetration boundary conditions along it,
\begin{equation}
\left(\tilde{\boldsymbol{u}}_{\Vert},\tilde{u}_{\bot}\right)|_{\tilde{z}=0}=\left(\tilde{\boldsymbol{u}}_{\mathrm{slip}}(\tilde{x},\tilde{y},\tilde{t}),0\right),\quad\left(\tilde{\boldsymbol{u}}_{\Vert},\tilde{u}_{\bot}\right)|_{\tilde{z}=\tilde{h}}=\left(\boldsymbol{0},\frac{\partial\tilde{d}}{\partial\tilde{t}}\right).\label{BC Flow}
\end{equation}
In this work, we consider a non-uniform slip velocity
and  specifically focus on the electro-osmotic slip which arises over
electrically charged surfaces due to interaction of an externally
applied electric field with the excess of net charge in the electric
double layer. In the thin-double-layer limit, such interaction results in bulk fluid motion outside
the outer edge of the electric double layer according to the Helmholtz\textendash Smoluchowski equation \citep{hunter2001foundations},
\begin{equation}
\tilde{\boldsymbol{u}}_{\mathrm{slip}}=-\frac{\tilde{\varepsilon}\tilde{\zeta}(\tilde{x},\tilde{y})}{\tilde{\mu}}\tilde{\boldsymbol{E}},\label{Slip}
\end{equation}
where $\tilde{\varepsilon}$ is the fluid permittivity, $\tilde{\zeta}$
is the zeta potential on the bottom surface and $\tilde{\boldsymbol{E}}$
is the tangential imposed electric field. 

\subsection{Scaling analysis and non-dimensionalization}

Scaling by the characteristic dimensions, we define the following normalized quantities: $(x,y,z)=(\tilde{x}/\tilde{l}^{*},\tilde{y}/\tilde{l}^{*},\tilde{z}/\tilde{h}_{0})$, $(u,v,w)=(\tilde{u}/\tilde{u}^{*},\tilde{v}/\tilde{u}^{*},\tilde{w}/\tilde{w}^{*})$, $p=\tilde{p}/\tilde{p}^{*}$, 
$t=\tilde{t}/\tilde{t}^{*}$, $d=\tilde{d}/\tilde{d}^{*}$ and $h=\tilde{h}/\tilde{h}_{0}$. As noted by \citet{peng2015displacement},
the bending-tension length scale $\tilde{l}_{BT}=\sqrt{\tilde{B}/\tilde{T}}$
determines the relative importance of bending and tension forces in
the elastic response of the sheet; for $\tilde{l}^{*}\ll\tilde{l}_{BT}$
bending forces dominate, whereas for $\tilde{l}^{*}\gg\tilde{l}_{BT}$
tension forces dominate. A convenient dimensionless number when scaling (\ref{Elastic balance}) is $\lambda$, defined as
\begin{equation}
\lambda\equiv\frac{\mathrm{Bending}}{\mathrm{Stretching}}=\left(\frac{\tilde{l}_{BT}}{\tilde{l}^{*}}\right)^{2}=\frac{\tilde{B}}{\tilde{T}\tilde{l}^{*2}}=\frac{1}{12(1-\nu^{2})}\frac{\tilde{E}_{\mathrm{Y}}\tilde{h}_{\mathrm{m}}}{\tilde{T}}\left(\frac{\tilde{h}_{\mathrm{m}}}{\tilde{l}^{*}}\right)^{2}.\label{lamda text}
\end{equation}
In this study, our main focus is on deformations
which are comparable to, or larger than the sheet thickness, and
therefore the appropriate scaling for deformation is based on tension and not
on bending
\begin{equation}
\tilde{d}^{*}=\frac{\tilde{p}^{*}\tilde{l}^{*2}}{\tilde{T}},\label{N elastic balance}
\end{equation}
with which the elastic equation (\ref{Elastic balance}) takes the
form 
\begin{equation}
p=\lambda\nabla_{\Vert}^{4}d-\nabla_{\Vert}^{2}d+p_{\mathrm{e}}.\label{Norm elastic balance}
\end{equation}
From order-of-magnitude analysis of the continuity
and in-plane momentum equations (\ref{Continuity+Momentum}), it follows
that the perpendicular velocity and the pressure scale as $\tilde{w}^{*}\sim\epsilon\tilde{u}^{*}$
and $\tilde{p}^{*}\sim12\tilde{\mu}\tilde{u}^{*}/\epsilon^{2}\tilde{l}^{*}$,  respectively.
Using the kinematic boundary condition (\ref{BC Flow})
we obtain the scaling for the viscous\textendash elastic timescale $\tilde{t}^{*}\sim\tilde{d}^{*}/\tilde{w}^{*}\sim\tilde{d}^{*}/\epsilon\tilde{u}^{*}$.
Owing to a linear relation between $\tilde{d}^{*}$ and $\tilde{p}^{*}$
as well as between $\tilde{p}^{*}$ and $\tilde{u}^{*}$, the ratio
$\tilde{d}^{*}/\tilde{u}^{*}$ is not dependent on actuation force,
nor on $\tilde{\boldsymbol{u}}_{\mathrm{slip}}$ or $\tilde{p}_{\mathrm{e}}$. Thus,
the viscous\textendash elastic timescale solely depends on the properties of the fluid and the elastic medium, specifically the ratio $\tilde{\mu}/\tilde{T}$ and the geometry:
\begin{equation}
\tilde{t}^{*}=\frac{12\tilde{\mu}\tilde{l}^{*4}}{\tilde{T}\tilde{h}_{0}^{3}}=\frac{12\tilde{\mu}\tilde{l}^{*}}{\epsilon^{3}\tilde{T}}.\label{Viscous-elastic time scale}
\end{equation}
Based on this scaling, and following standard lubrication theory, from (\ref{Continuity+Momentum}), (\ref{BC Flow}) and (\ref{Viscous-elastic time scale}) we obtain the normalized Reynolds equation \citep[p. 313] {Leal}
\begin{equation}
\frac{\partial d}{\partial t}-\boldsymbol{\nabla}_{\Vert}\cdot[h^{3}\boldsymbol{\nabla}_{\Vert}p]=-\frac{1}{2}\boldsymbol{\nabla}_{\Vert}\cdot [h \boldsymbol{u}_{\mathrm{slip}}].\label{Evoluation calculated}
\end{equation}
where using the definition (\ref{Shallow}\textit{d}), the fluid thickness $h$ can be expressed as $ h=1+\beta d$.

\subsection{Viscous\textendash elastic governing equations for a pre-stretched elastic sheet }
Substituting (\ref{Norm elastic balance}) into (\ref{Evoluation calculated}), yields the nonlinear governing equation
\begin{equation}
\frac{\partial d}{\partial t}-\boldsymbol{\nabla}_{\Vert}\cdot[h^{3}\boldsymbol{\nabla}_{\Vert}(\lambda\nabla_{\Vert}^{4}d-\nabla_{\Vert}^{2}d)]=-\boldsymbol{\nabla}_{\Vert}\cdot [h \boldsymbol{f}_{F}]+\boldsymbol{\nabla}_{\Vert}\cdot[h^{3}\boldsymbol{\nabla}_{\Vert}f_{E}],\label{Evoluation calculated-1}
\end{equation}
where we define $\boldsymbol{f}_{F}=f_{Fx}\boldsymbol{\hat{x}}+f_{Fy}\boldsymbol{\hat{y}}=\boldsymbol{u}_{\mathrm{slip}}/2$
and $f_{E}=p_{\mathrm{e}}$ as the actuation mechanisms. The subscript $F$ refers to driving force applied to the sheet due to the non-uniform slip velocity  in the fluid and the subscript $E$ refers to driving force due to the pressure applied directly to the elastic sheet.

Assuming small elastic deformations, $\tilde{d}^{*}\ll\tilde{h}_{0}$, yields the linearized viscous\textendash elastic governing equation in terms of deformation accounting for both bending and pre-stretching \citep{kodio2016lubricated}, and containing a source term that depends on the external forces
\begin{equation}
\frac{\partial d}{\partial t}-\lambda\mathbf{\mathbf{\nabla_{\Vert}^{\mathrm{6}}}}d+\mathbf{\mathbf{\nabla_{\Vert}^{\mathrm{4}}}}d=-\boldsymbol{\nabla}_{\Vert}\cdot\boldsymbol{f}_{F}+\mathbf{\mathbf{\nabla_{\Vert}^{\mathrm{2}}}}f_{E}.\label{Governing equation deformation}
\end{equation}
For completeness, the corresponding dimensional equation is given by
\begin{equation}
\frac{\partial\tilde{d}}{\partial\tilde{t}}-\frac{\tilde{h}_{0}^{3}}{12\tilde{\mu}}\tilde{\boldsymbol{\nabla}}_{\Vert}\cdot[\tilde{\boldsymbol{\nabla}}_{\Vert}(\tilde{B}\tilde{\nabla}_{\Vert}^{4}\tilde{d}-\tilde{T}\tilde{\nabla}_{\Vert}^{2}\tilde{d})]=-\tilde{h}_{0}\tilde{\boldsymbol{\nabla}}_{\Vert}\cdot\tilde{\boldsymbol{f}}_{F}+\frac{\tilde{h}_{0}^{3}}{12\tilde{\mu}}\mathbf{\mathbf{\tilde{\nabla}_{\Vert}^{\mathrm{2}}}}\tilde{f}_{E}.\label{Dimensional}
\end{equation}
The governing equations (\ref{Evoluation calculated-1})--(\ref{Dimensional}) are supplemented by the boundary conditions 
\refstepcounter{equation}
$$
 d=0,\quad\frac{\partial^{2}d}{\partial x^{2}}=0,\quad\frac{\partial^{4}d}{\partial x^{4}}=0\quad\mathrm{at}\quad x=0,\,l_{\mathrm{m}}\quad \mathrm{and}\quad0\leq y\leq w_{\mathrm{m}},
  \eqno{(\theequation{a-c})}\label{d BC 1}
$$
\refstepcounter{equation}
$$
 d=0,\quad\frac{\partial^{2}d}{\partial y^{2}}=0,\quad\frac{\partial^{4}d}{\partial y^{4}}=0\quad\mathrm{at}\quad y=0,\,w_{\mathrm{m}}\quad \mathrm{and}\quad0\leq x\leq l_{\mathrm{m}},
  \eqno{(\theequation{a-c})}\label{d BC 2}
$$ and the initial condition $d(x,y,t=0)=0$. The first two boundary conditions (\ref{d BC 1}\textit{a\textendash b}) and (\ref{d BC 2}\textit{a\textendash b})
correspond to no deflection and no moment at the boundaries, whereas
the last conditions (\ref{d BC 1}\textit{c}) and (\ref{d BC 2}\textit{c}) are obtained from (\ref{Elastic balance}) by further
assuming that the fluidic and external pressures
are both zero at the boundaries, $p=p_{\mathrm{e}}=0$ \citep[see, e.g.,][]{kodio2016lubricated}. We
note that the condition $p=0$ is motivated by the experimental setup
shown in figure \ref{F1}, where the fabricated chamber has open boundaries, which are well represented by zero gauge pressure. In addition, for the case of a tension-dominant regime, which is the main focus of this work, setting $\lambda=0$ in  (\ref{Evoluation calculated-1})--(\ref{Dimensional})
yields fourth-order governing equations. The boundary conditions (\ref{d BC 1}\textit{a--b}) and (\ref{d BC 2}\textit{a--b}) hold for this case, with the second derivative requirement arising directly from the zero pressure condition (rather than from the no-moment requirement in the $6^{\mathrm{th}}$ order case).

The governing equation (\ref{Governing equation deformation}) can be solved by several methods. For the boundary conditions under consideration (\ref{d BC 1})--(\ref{d BC 2}), eigenfunction expansions or a Green's functions approach are particularly suitable. We refer the reader to \citet{prosperetti2011advanced} for both methods of solution, and note that in practice these calculations may be cumbersome. The general solution based on a Green's function approach for the case presented here is provided in appendix B of the supplementary material.  However, we stress that both eigenfunctions and Green's functions approaches would be far more complex to implement if boundary conditions other than those presented here are adopted. Furthermore, analytical treatment would be greatly complicated, if at all possible, in non-rectangular and non-circular domains of interest.

\section{Deformations due to a square-shaped actuation region}
We now consider the non-uniform slip velocity (\ref{Slip}) as a driving force and specifically focus on a square-shaped actuation of the form
\begin{equation}
f_{F_{x}}(x,y,t)=\left\{\begin{array}{ll}
\hspace{-1mm}\displaystyle{E(t)} &  |x-c_{x}|\leq L\quad\mathrm{and}\quad|y-c_{y}|\leq L\vspace{0.5mm}\\
\displaystyle{0} & \mathrm{otherwise},
\end{array}\label{Square-shaped actuation} \right.
\end{equation}corresponding to a square-shaped zeta-potential distribution, where
$E(t)$ is a spatially homogeneous and time-dependent electric field along the $\boldsymbol{\hat{x}}$ axis. Here $2L$ is the side length of the actuation
square, whereas $c_{x}$ and $c_{y}$ indicate the $x$- and $y$-coordinates
of its center. For convenience, hereafter we set $l_{\mathrm{m}}=w_{\mathrm{m}}=\pi$.

For a suddenly applied actuation, $E(t)=H(t)$, where $H$ is the Heaviside step function, using closed-form solutions derived in appendix B of the supplementary material, we obtain the deformation field resulting from (\ref{Square-shaped actuation})
\begin{subequations}
\begin{equation}
d(x,y,t)=\frac{16}{\pi^{2}}\sum_{m,\,n=1}^{\infty}\mathcal{A}(m,n;L,c_{x},c_{y})\frac{\sin\left(mx\right)\sin\left(ny\right)}{nF(m,n;\lambda)}\left(1-\mathrm{e}^{-F(m,n;\lambda)t}\right),\label{d square Ff}
\end{equation}
\begin{equation}
\mathcal{A}=\cos(mc_{x})\sin(nc_{y})\sin(mL)\sin(nL),\quad F=\left(m^{2}+n^{2}\right)^{2}\left[\lambda\left(m^{2}+n^{2}\right)+1\right].\label{A2}
\end{equation}
\end{subequations}
In appendix C of the supplementary material, we use numerical computations to verify the analytical solution  (\ref{d square Ff}) for the case of $\lambda=0$, showing excellent agreement.

\begin{figure}
 \centerline{\includegraphics[scale=1]{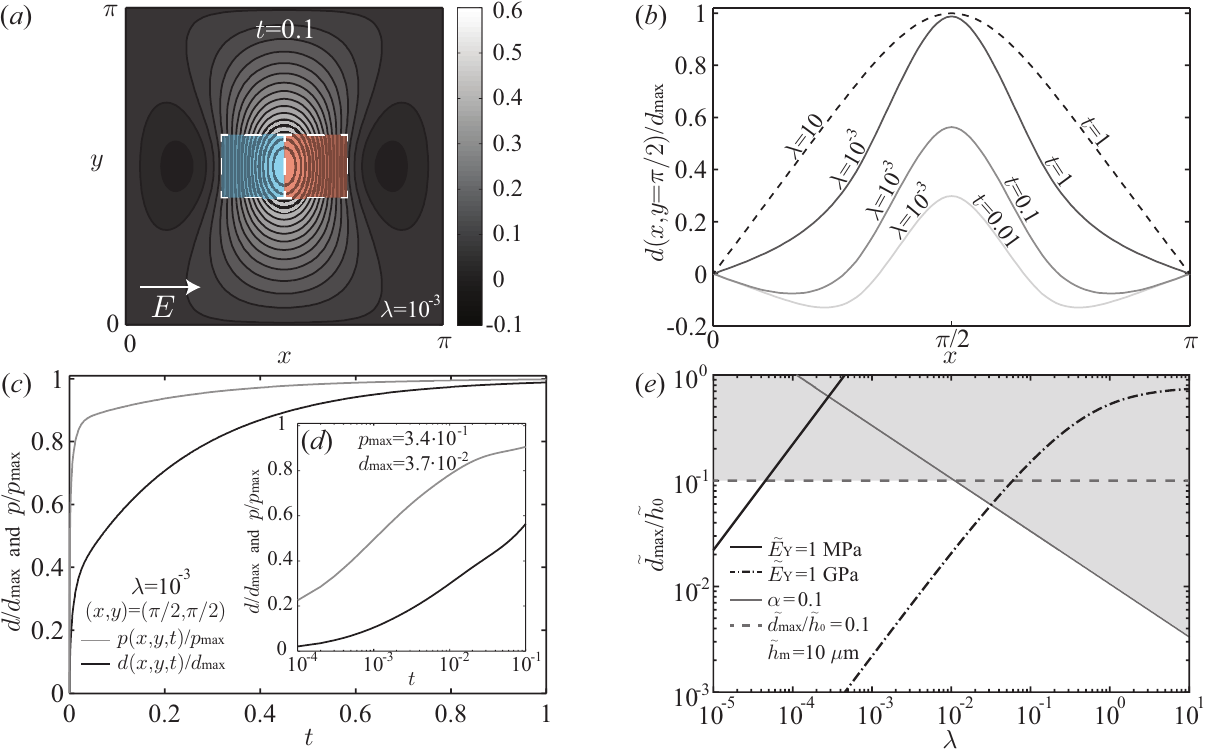}}
\caption{Investigation of the deformation field resulting from two square-shaped regions with opposite signs of zeta potential subjected to a constant electric field suddenly applied at $t=0$. (a) The deformation field (colormap) at $t=0.1$, superposed on a schematic illustration of the two oppositely directed electro-osmotic actuation regions. 
(b) The scaled steady-state deformation $d/d_\mathrm{max}$ along the $\boldsymbol{\hat{x}}$ axis, showing the evolution of the deformation for $\lambda=10^{-3}$, and a comparison between the steady-state deformation at $\lambda=10^{-3}$ (tension dominant, with $d_\mathrm{max}=3.7\cdot10^{-2}$) and $\lambda=10$ (bending dominant, with $d_\mathrm{max}=1.2\cdot10^{-3}$). (c--d) The evolution of the maximum deformation and pressure as a function of time. (e) The maximum deformation $\tilde{d}_\mathrm{max}$ at steady state, scaled by $\tilde{h}_{0}$, as a function of $\lambda$, for $\tilde{E}_{\mathrm{Y}}=1\,\mathrm{MPa}$ and $\tilde{E}_{\mathrm{Y}}=1\,\mathrm{GPa}$. The dashed and solid gray lines indicate the validity boundaries $\tilde{d}_\mathrm{max}/\tilde{h}_{0}=0.1$ and $\alpha=0.1$, respectively, where the grayed out region is beyond the validity of the model. All calculations were performed using $\tilde{l}^{*}=5$ mm, $\tilde{h}_{0}=100\,\mu\mathrm{m}$,
$\tilde{p}^{*}\approx1$ Pa, $\tilde{h}_{\mathrm{m}}=10\,\mu\mathrm{m}$, $\nu=0.5$, $c_{x,1}=2\pi/5$, $c_{x,2}=3\pi/5$, $c_{y,1}=c_{y,2}=\pi/2$ and $L=\pi/10$.}
\label{F3}
\end{figure}

Figure \ref{F3}(a) presents the deformation field (\ref{d square Ff}) at $t=0.1$, resulting from two square-shaped regions with opposite signs of zeta potential, subjected to an electric field suddenly applied at $t=0$, (see also supplementary information video 1). The opposing flows result in a positive internal pressure at the interface between the two regions, and negative pressure at the far edges of the squares.
Figure \ref{F3}(b) presents the deformation field (along the $\boldsymbol{\hat{x}}$ axis) as a function of time, for a pre-stretching-dominant regime $\lambda=10^{-3}$. The steady-state deformation ($t=1$) with $d_\mathrm{max}=3.7\cdot10^{-2}$ is compared with that obtained for a bending-dominant regime ($\lambda=10$) having $d_\mathrm{max}=1.2\cdot10^{-3}$, showing that the higher order of the dominant term in the bending regime effectively results in additional averaging of the fluidic pressure and thus reduces the spatial resolution and the magnitude of the deformation field. Figures \ref{F3}(c--d) present the development of the maximum pressure and deformation as a function of time towards steady state, where $p_\mathrm{max}=3.4\cdot10^{-1}$ and $d_\mathrm{max}=3.7\cdot10^{-2}$. Importantly, while for a rigid configuration actuation of the electric field would result in an instantaneous jump in pressure (on an acoustic timescale), in the elastic case the evolution of the pressure is coupled to that of the deformation, as given by (\ref{Evoluation calculated}). Nevertheless, for the case presented, the rise time to 50$\%$ of the final pressure is $t=10^{-3}$, nearly two orders of magnitude shorter than the viscous\textendash elastic timescale $t=0.7\cdot10^{-1}$ in which the deformation reaches 50$\%$ of its steady-state value. This time delay between pressure and deformation is also evident in periodic actuations (see supplementary information video 2) where the deformation phase lags behind that of the pressure. We note that at extremely short timescales, inertial effects must be included to properly describe the deformation dynamics. Figure \ref{F3}(e) presents the maximum deformation (scaled by $\tilde{h}_{0}$) at steady state as a function of $\lambda$, for a fixed sheet thickness and two different elastic moduli. Consistent with the scaling (\ref{N elastic balance})
and non-dimensional solution (\ref{d square Ff}), as $\lambda$ increases ($\tilde{T}$ decreases), the magnitude of the deformation increases until it reaches a constant value when the problem is dominated entirely by bending. The dashed and solid gray lines respectively indicate the range of validity of the model, where the deformation is no longer small, $\tilde{d}_\mathrm{max}/\tilde{h}_{0}=0.1$, and where internal stretching is non negligible, $\alpha=0.1$. For clarity, the regions of the graph which are beyond the validity of the model are grayed out.

\section{Effect of pre-stretching on the resolution, magnitude and timescale of the deformation field} 
As we are aiming to utilize fluidic actuation as a mechanism to create desired deformation in the elastic sheet, it is of interest to examine the tradeoff between the magnitude and timescale of deformations and the attainable resolution, for a given amplitude of the actuation force. 
Any desired deformation can be written as a Fourier series on a finite domain, where the resulting spectrum of frequencies would depend on the desired shape and on its position relative to the boundaries. 
Consider a steady-state deformation of the form, 
\begin{equation}
\tilde{d}\left(\tilde{x},\tilde{y}\right)=\tilde{d}_{0}\sin\left(k_{x}\pi\frac{\tilde{x}}{\tilde{l}_{\mathrm{m}}}\right)\sin\left(k_{y}\pi\frac{\tilde{y}}{\tilde{w}_{\mathrm{m}}}\right),\label{sine deformation}
\end{equation}
created due to a non-uniform Helmholtz\textendash Smoluchowski slip velocity, (\ref{Slip}), acting in the $\boldsymbol{\hat{x}}$ direction. Here $k_{x}$ and
$k_{y}$ are wavenumbers in the $\boldsymbol{\hat{x}}$ and $\boldsymbol{\hat{y}}$
directions, respectively, and $\tilde{d}_{0}$ is the amplitude yet to be determined. Using (\ref{Slip}) and (\ref{Dimensional}), the zeta potential required for generating
the deformation (\ref{sine deformation}) is given by
\begin{equation}
\tilde{\zeta}(\tilde{x},\tilde{y})=\frac{\tilde{h}_{0}^{2}}{6\tilde{\varepsilon}\tilde{E}}\int\left(-\tilde{B}\tilde{\nabla}_{\Vert}^{6}\tilde{d}+\tilde{T}\tilde{\nabla}_{\Vert}^{4}\tilde{d}\right)\mathrm{d}\tilde{x}=\tilde{\zeta}_{0}\cos\left(k_{x}\pi\frac{\tilde{x}}{\tilde{l}_{\mathrm{m}}}\right)\sin\left(k_{y}\pi\frac{\tilde{y}}{\tilde{w}_{\mathrm{m}}}\right),\label{zeta for sine}
\end{equation}
enabling us to explicitly express the amplitude $\tilde{d}_{0}$ in terms of relevant
physical quantities, 
\begin{equation}
\tilde{d}_{0}=-\frac{\tilde{6\varepsilon}\tilde{\zeta}_{0}\tilde{E}}{\pi^{3}\tilde{h}_{0}^{2}}\frac{k_{x}/\tilde{l}_{\mathrm{m}}}{\left((k_{x}/\tilde{l}_{\mathrm{m}})^{2}+(k_{y}/\tilde{w}_{\mathrm{m}})^{2}\right)^{2}\left(\pi^{2}\tilde{B}\left[(k_{x}/\tilde{l}_{\mathrm{m}})^{2}+(k_{y}/\tilde{w}_{\mathrm{m}})^{2}\right]+\tilde{T}\right)}.\label{Amp of deformation}
\end{equation}
Substituting (\ref{zeta for sine}) into (\ref{Dimensional}), we obtain the corresponding time-dependent solution
\begin{equation}
\tilde{d}\left(\tilde{x},\tilde{y},\tilde{t}\right)=\tilde{d}_{0}\sin\left(k_{x}\pi\frac{\tilde{x}}{\tilde{l}_{\mathrm{m}}}\right)\sin\left(k_{y}\pi\frac{\tilde{y}}{\tilde{w}_{\mathrm{m}}}\right)\left(1-\mathrm{e}^{-\tilde{t}/\tilde{\tau}_{0}}\right),\label{Dimensional sine evolution}
\end{equation}
that evolves towards steady-state deformation $\tilde{d}\left(\tilde{x},\tilde{y}\right)$,
(\ref{sine deformation}), and provide the viscous\textendash elastic timescale
$\tilde{\tau}_{0}$ required to achieve this steady state, 
\begin{equation}
\tilde{\tau}_{0}=\frac{12\tilde{\mu}}{\pi^{4}\tilde{h}_{0}^{3}}\frac{1}{\left((k_{x}/\tilde{l}_{\mathrm{m}})^{2}+(k_{y}/\tilde{w}_{\mathrm{m}})^{2}\right)^{2}\left(\pi^{2}\tilde{B}\left[(k_{x}/\tilde{l}_{\mathrm{m}})^{2}+(k_{y}/\tilde{w}_{\mathrm{m}})^{2}\right]+\tilde{T}\right)}.\label{Timescale tau0}
\end{equation}
\begin{figure}
 \centerline{\includegraphics{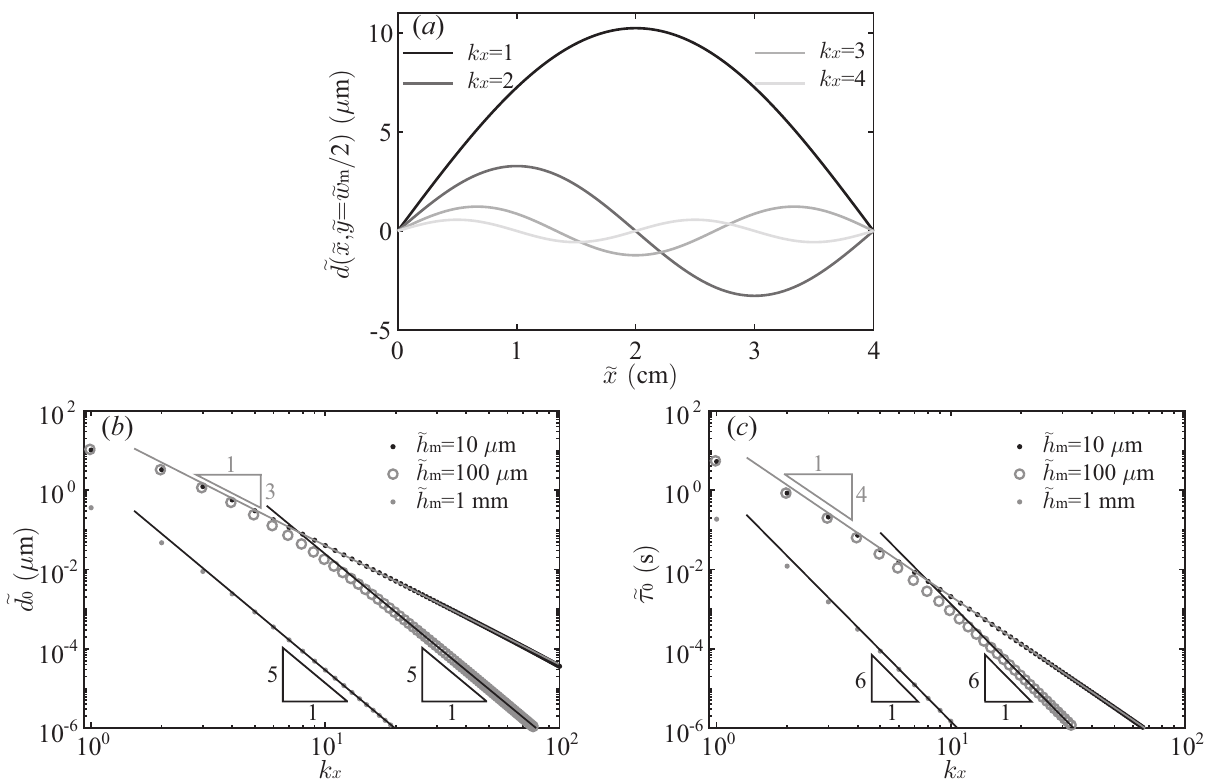}}
\caption{The effect of wavenumber on the magnitude of the deformation and the timescale of viscous\textendash elastic interaction. (a) Deformation in dimensional form along the $\boldsymbol{\hat{x}}$ axis for various values of $k_{x}$, with $k_{y}=1$ and $\tilde{h}_{\mathrm{m}}=10$ $\mu$m. (b--c) The maximum (dimensional) deformation $\tilde{d}_{0}$ and timescale $\tilde{\tau}_{0}$, respectively, as a function of wavenumber $k_{x}$, for $k_{y}=1$ and different values of the elastic sheet thickness, $\tilde{h}_{\mathrm{m}}$. All calculations were performed using non-uniform electro-osmotic actuation as the driving mechanism, with $\tilde{h}_{0}=100\,\mu$m, $\tilde{l}_{\mathrm{m}}=\tilde{w}_{\mathrm{m}}$= 4 cm, $\tilde{E}_{\mathrm{Y}}=0.3$ GPa, $\nu=0.5$, $\tilde{T}=15 $ $\mathrm{Pa\,m}$, $\tilde{\zeta}_{0}=-70$ mV, $\tilde{E}=$100 $\mathrm{V\,cm^{-1}}$ and $\tilde{\varepsilon}=7.08\cdot10^{-10}\,\mathrm{F\,m^{-1}}$.}
\label{F4}
\end{figure}

Using the physical values noted in its caption, figure \ref{F4}(a) presents
the resulting steady-state deformation (\ref{sine deformation}) along
the $\boldsymbol{\hat{x}}$ axis for various values of $k_{x}$, with
$k_{y}=1$ and $\tilde{h}_{\mathrm{m}}=10\,\mu\mathrm{m}$, and clearly shows
the reduction in deformation magnitude as the wavenumber increases.
Furthermore, (\ref{Amp of deformation}) and (\ref{Timescale tau0})
indicate that the scaling of the amplitude $\tilde{d}_{0}$ and timescale $\tilde{\tau}_{0}$ with the wavenumber depends on the relative
contribution of the bending and tension terms in the denominator.
When $\pi^{2}\tilde{B}k_{x}^{2}\ll\tilde{T}\tilde{l}_{\mathrm{m}}^{2}$, pre-stretching
is dominant over bending forces and the deformation $\tilde{d}_{0}$
and timescale $\tilde{\tau}_{0}$ scale as $k_{x}^{-3}$ and $k_{x}^{-4}$,
respectively. However, when $\pi^{2}\tilde{B}k_{x}^{2}\gg\tilde{T}\tilde{l}_{\mathrm{m}}^{2}$,
bending is dominant, and the deformation and timescale scale as
$k_{x}^{-5}$ and $k_{x}^{-6}$, respectively. From the definition
of $\tilde{B}$, the condition for pre-stretching dominance can be
expressed as $\pi^{2}\tilde{E}_{\mathrm{Y}}\tilde{h}_{\mathrm{m}}^{3}k_{x}^{2}/12(1-\nu^{2})\tilde{T}\tilde{l}_{\mathrm{m}}^{2}\ll1$.
Figures \ref{F4}(b--c) present the maximum deformation and the timescale as
a function of $k_{x}$, respectively, for three values of the membrane
thickness, $\tilde{h}_{\mathrm{m}}$, showing that the deformation of a 10 $\mu$m
thick sheet scales as $k_{x}^{-3}$ throughout the investigated range,
whereas the timescale scales as $k_{x}^{-4}$. As the membrane thickness
increases, the bending effect becomes apparent and for $\tilde{h}_{\mathrm{m}}=100\,\mu\mathrm{m}$
the amplitude and the timescale scale as $k_{x}^{-3}$ and $k_{x}^{-4}$
for sufficiently low $k_{x}$ values, but settle on a $k_{x}^{-5}$
and $k_{x}^{-6}$ dependence for high wavenumbers. For $\tilde{h}_{\mathrm{m}}=1\,\mathrm{m}\mathrm{m}$,
the bending effect is dominant and thus the amplitude and the timescale
scale as $k_{x}^{-5}$ and $k_{x}^{-6}$ even for low wavenumbers. We note that for the set of parameters chosen here, even the first modes corresponding to the largest deformation satisfy the small-deformation requirement of the model.

\section{Effect of discretized actuation on the deformation field}

Equation (\ref{Governing equation deformation}) may be solved to obtain the actuation field required to achieve pre-defined deformation patterns of elastic plates, which may be desired in engineering applications. However, implementation of the resulting continuous actuation distribution may be challenging in practice. 
We here consider a discrete actuation profile, consisting of $D\times D$ individual squares, and seek to minimize the error between the resulting and desired deformation, for a given level of discretization. Specifically, actuating square $i$ with amplitude $a_{i}$ yields
a deflection $a_{i}d_{i}(x,y)$ given by (\ref{d square Ff}), such that the resulting deformation is $\sum_{i=1}^{D^2} a_{i}d_{i}(x,y)$.
For a given desired deformation $d(x,y)$, we follow a least-squares
method \citep{bjorck1996numerical} and seek to minimize the error
\begin{equation}
\int_{0}^{l_{\mathrm{m}}}\int_{0}^{w_{\mathrm{m}}}\left(\sum_{{i,\,j=1}}^{D^2}a_{i}d_{i}-d\right)^{2}\mathrm{d}x\mathrm{d}y=\sum_{{i,\,j=1}}^{D^2}a_{i}a_{j}\mathsfbi{A}_{i,j}-2\sum_{{i=1}}^{D^2}a_{i}b_{i}+c,\label{Discret 1}
\end{equation}
where 
\begin{equation}
\mathsfbi{A}_{i,j}=\int_{0}^{l_{\mathrm{m}}}\int_{0}^{w_{\mathrm{m}}}d_{i}d_{j}\mathrm{d}x\mathrm{d}y,\quad b_{i}=\int_{0}^{l_{\mathrm{m}}}\int_{0}^{w_{\mathrm{m}}}d_{i}d\mathrm{d}x\mathrm{d}y,\quad c=\int_{0}^{l_{\mathrm{m}}}\int_{0}^{w_{\mathrm{m}}}d^{2}\mathrm{d}x\mathrm{d}y.\label{Discret 2}
\end{equation}
Differentiating (\ref{Discret 2}) with respect to each $a_{i}$ and
equating to zero yields the optimality condition
\begin{equation}
\mathsfbi{A}\boldsymbol{a}=\boldsymbol{b},\label{Discret 3}
\end{equation}
where $\boldsymbol{a}=[a_{1},a_{2},...,a_{D^{2}}]^{T}$, $\boldsymbol{b}=[b_{1},b_{2},...,b_{D^{2}}]^{T}$ 
and $\mathsfbi{A}$ is a symmetric $D^{2}\times D^{2}$ matrix with rank $D^{2}-D$
and thus is not invertible (singular). The singularity of $\mathsfbi{A}$ stems
from the fact that the source term in (\ref{Governing equation deformation})
depends on gradients of the actuation field, thus allowing an associated
gauge freedom in the choice of actuation (coefficients $a_{i}$) without
modifying the resulting deformation. Specifically, for the case of
a driving force $f_{F_{x}}$ acting in $\boldsymbol{\hat{x}}$ the
direction, it follows that adding an arbitrary function $f_{0}(y)$
to the driving force, which has $D$ values in discrete form, will not
modify the resulting deformation. 

To determine the coefficients $a_{i}$, we first reduced the matrix
$A$ and vector $\boldsymbol{b}$ only to rows with corresponding
non-zero eigenvalues of $\mathsfbi{A}$, obtaining a $(D^{2}-D)\times (D^{2})$ matrix
and a $(D^{2}-D)\times (1)$ vector, respectively. We then solved the reduced
system (\ref{Discret 3}) and found the vector $\boldsymbol{a}$
using MATLAB's routine \texttt{lsqminnorm} (release R2017b, Mathworks, USA),
which computes the minimum least-squares solution of the system.
\begin{figure}
 \centerline{\includegraphics[scale=1]{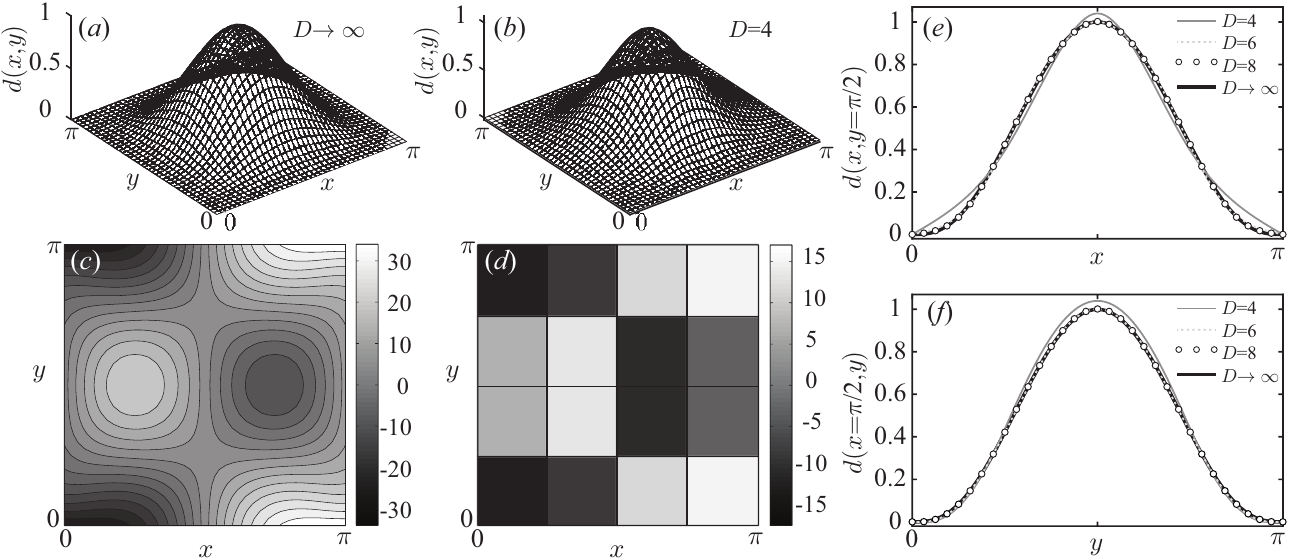}}
\caption{The effect of actuation discretization on the shape and the magnitude
of the deformation. (a) Deformation formed by the continuous driving
force $f_{F_{x}}$ shown in (c), acting in the $\boldsymbol{\hat{x}}$
direction. (b) Deformation obtained from $4\times 4$ square-shaped actuation regions, each having a uniform value which is found
by solving the least-squares problem (\ref{Discret 3}) and shown
in (d). (e--f) Deformation along the $\boldsymbol{\hat{x}}$ and $\boldsymbol{\hat{y}}$
axes, respectively, for different levels of discretization.}
\label{F5}
\end{figure}
As an illustrative example, we consider a localized deformation,
\begin{equation}
d(x,y)=\left(2/\pi\right)^{12}x^{3}\left(\pi-x\right)^{3}y^{3}\left(\pi-y\right)^{3},\label{4}
\end{equation}
shown in figure \ref{F5}(a), and for simplicity assume a membrane regime
(dominant pre-stretching, $\lambda=0$). Using (\ref{Governing equation deformation}),
the continuous driving force $f_{F_{x}}$ required to create this
deformation (at steady state) is given by
\begin{equation}
f_{F_{x}}(x,y)=-\int_{\pi/2}^{x}\mathbf{\mathbf{\nabla_{\Vert}^{\mathrm{4}}}}d(x',y)\mathrm{d}x',\label{5}
\end{equation}
and shown in figure \ref{F5}(c). Figure \ref{F5}(d) shows the discretization of
this field into $4\times 4$ square-shaped regions, each assigned with a uniform
value obtained from the least-squares method solution of (\ref{Discret 3}).
Figure \ref{F5}(b) presents the resulting deformation obtained by superposition
of solutions for individual squares (\ref{d square Ff}), using the
values for forcing shown in figure \ref{F5}(d). Figures \ref{F5}(e--f) present the
deformation along the $\boldsymbol{\hat{x}}$ and $\boldsymbol{\hat{y}}$
axes, respectively, for different levels of discretization. While,
as expected, discretization results in undesired oscillations of the
deformation field due to high wavenumbers which are unbalanced, even
a relatively coarse discretization with $D=4$ allows maintaining
a fairly localized deformation, with amplitude only larger by $4\%$
than the desired one. For
higher levels of discretization ($D=6$ and $D=8$), the difference
between the resulting and desired deformations is almost indistinguishable.
\begin{figure}
 \centerline{\includegraphics[scale=1]{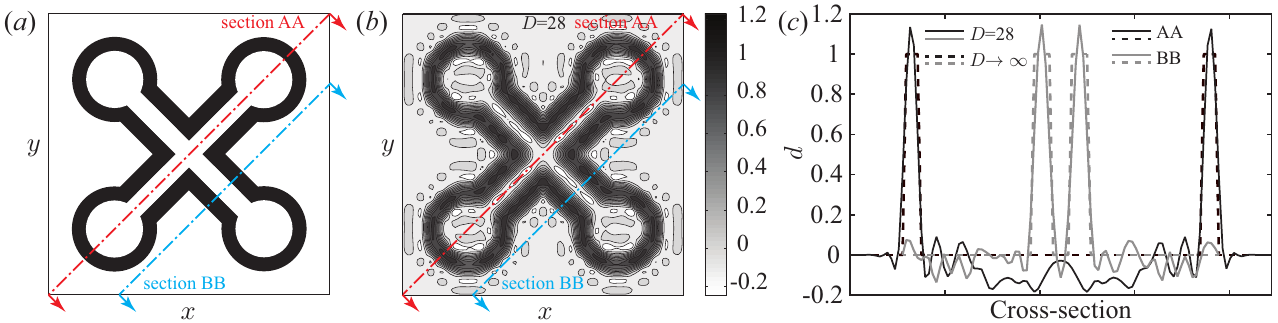}}
\caption{Demonstration of the use of a least-squares-discretization
method for creation of a complex deformation. (a) Top view of a microfluidic
configuration, in which black lines represent the confining walls,
characterized by step-like behavior. (b) The deformation field (colourmap)
obtained from $28\times 28$ square-shaped actuation regions, each having a
uniform value found by solving the least-squares problem
(\ref{Discret 3}). (c) Deformation along the two cross-sections,
depicted in (a) and (b), showing that discretization results in undesired
oscillations of the deformation field. Dashed and solid lines correspond
to the chosen and obtained deformation, respectively.}
\label{F6}
\end{figure}

To challenge our method of solution, we consider
the microfluidic configuration shown in figure \ref{F6}(a), in which the deformation magnitude is dictated strictly by 0 and 1, represented in the figure
by white and black colors, respectively. Clearly, the exact description of
this deformation field requires an infinite number of wavenumbers which
cannot be represented by a finite grid discretization. Nevertheless,
the deformation obtained using $28\times 28$ square-shaped actuation regions
is able to reproduce the main features of the geometry as shown in
figure \ref{F6}(b), though with undesired oscillations. For further clarification, figure \ref{F6}(c) presents the deformation
along the two cross-sections, illustrated in figures \ref{F6}(a--b), showing
that the maximal and minimal deformations are approximately by $11\%$
and $18\%$ larger than the desired one, respectively.

\section{Investigation of nonlinear effects} 
In the previous sections we restricted our analysis
to a strongly pre-stretched elastic sheet, $\alpha\ll1$, and small
elastic deformations, $\beta\ll1$, and solved the linear governing
equation (\ref{Governing equation deformation}) for the deformation
field. In this section, we relax these restrictions and explore the
nonlinear effects arising from the hydrodynamic nonlinearity in the
Reynolds equation, $h^{3}$, and the nonlinear coupling between the
tension and deformation in the F$\mathrm{\ddot{o}}$ppl\textendash von K$\mathrm{\acute{a}rm\acute{a}n}$ eqautions. However, we limit our nonlinear analysis to positive deformations, $d>0$. We first
study theoretically the weakly nonlinear effect of internal tension
on the deformation ($\mathsection$ 6.2) and then investigate numerically
the nonlinear effects considering finite $\alpha$ and $\beta$ ($\mathsection$
6.3). To this end, we consider an axisymmetric geometry which allows
reduction to a single spatial variable and thus greatly simplifies
theoretical and numerical investigation. This simple case is useful in providing physical insight on the nonlinear effects
of $\alpha$ and $\beta$ on the deformation and tension field. We also note the analysis of the axisymmetric configurations is relevant to adaptive optics applications, in which the liquid within a circular microchamber, covered with a thin elastic membrane, is pressurized to deform the membrane forming a plano-convex microlens \citep{chronis2003tunable}.

\subsection{Coupled Reynolds and nonlinear F$\ddot{o}$ppl\textendash von K$\acute{a}rm\acute{a}n$ equations}

We consider a circular elastic sheet of radius $\tilde{R}_{\mathrm{m}}$
subjected to axisymmetric driving forces acting either through the pressure applied directly to the elastic sheet or through the non-uniform slip velocity in the fluid.
For simplicity, we neglect the effect of bending and
consider a membrane (tension-dominant) regime, $\lambda\ll1$.
Based on these assumptions, the deformations of the elastic membrane
are described by the axisymmetric F$\mathrm{\ddot{o}}$ppl\textendash von K$\mathrm{\acute{a}rm\acute{a}n}$ equations for the membrane  \citep[see][]{ landau1959,lister2013viscous,zheng2015propagation}
\begin{equation}
\tilde{p}=-\mathbf{\mathbf{\boldsymbol{\tilde{\nabla}}}}_{r}\cdot(\tilde{T}_{r}\mathbf{\mathbf{\boldsymbol{\tilde{\nabla}}}}_{r}\tilde{d})+\tilde{p}_{\mathrm{e}},\label{von-Karman stretching polar dim (a)}
\end{equation}
and
\begin{equation}
\frac{1}{\tilde{r}}\frac{\partial}{\partial\tilde{r}}\left(\tilde{r}^{3}\frac{\partial\tilde{T}_{r}}{\partial\tilde{r}}\right)=-\frac{\tilde{E}_{\mathrm{Y}}\tilde{h}_{\mathrm{m}}}{2}\left(\frac{\partial\tilde{d}}{\partial\tilde{r}}\right)^{2},\label{von-Karman stretching polar dim (b)}
\end{equation}
where $\mathbf{\mathbf{\boldsymbol{\tilde{\nabla}}}}_{r}=(\partial/\partial\tilde{r})\boldsymbol{\hat{r}}$
is the gradient in polar coordinates, $\tilde{r}=\sqrt{\tilde{x}^{2}+\tilde{y}^{2}}$
is the radial coordinate and $\tilde{T}_{r}(\tilde{r},\tilde{t})$
is the radial tension resulting from both external and internal tension
formed in the membrane.

We define the normalized radial coordinate $r=\tilde{r}/\tilde{l}$, radial tension $T_{r}=\tilde{T}_{r}/\tilde{T}$ and dimensionless
size of the membrane $R_{\mathrm{m}}=\tilde{R}_{\mathrm{m}}/\tilde{l}$. Substituting
the normalized variables into (\ref{von-Karman stretching polar dim (a)})--(\ref{von-Karman stretching polar dim (b)})
and using the results of scaling analysis in $\mathsection$ 2.1 we
obtain the non-dimensional F$\mathrm{\ddot{o}}$ppl\textendash von K$\mathrm{\acute{a}rm\acute{a}n}$ equations
\begin{equation}
p=-\boldsymbol{\nabla}_{r}\cdot(T_{r}\boldsymbol{\nabla}_{r}d)+p_{\mathrm{e}},\label{von-Karman stretching polar non-dim (a)}
\end{equation}
\begin{equation}
\frac{1}{r}\frac{\partial}{\partial r}\left(r^{3}\frac{\partial T_{r}}{\partial r}\right)=-\frac{1}{2}\alpha\left(\frac{\partial d}{\partial r}\right)^{2}.\label{von-Karman stretching polar non-dim (b)}
\end{equation}
Substituting (\ref{von-Karman stretching polar non-dim (a)}) into
the nonlinear Reynolds equation (\ref{Evoluation calculated}), we
obtain 
\begin{equation}
\frac{\partial d}{\partial t}+\boldsymbol{\nabla}_{r}\cdot[h^{3}\boldsymbol{\nabla}_{r}[\boldsymbol{\nabla}_{r}\cdot(T_{r}\boldsymbol{\nabla}_{r}d)]]=-\boldsymbol{\nabla}_{r}\cdot[h\boldsymbol{f}_{F}]+\boldsymbol{\nabla}_{r}\cdot[h^{3}\boldsymbol{\nabla}_{r}f_{E}].\label{Non-linear governing eqaution polar}
\end{equation}
The evolution equation (\ref{Non-linear governing eqaution polar}) and the second
F$\mathrm{\ddot{o}}$ppl\textendash von K$\mathrm{\acute{a}rm\acute{a}n}$ equation (\ref{von-Karman stretching polar non-dim (b)})
are two-way coupled nonlinear equations, governing the fluid\textendash structure
interaction, that should be solved at once to obtain both deformation
and tension fields. 

The governing equations (\ref{von-Karman stretching polar non-dim (b)}) and (\ref{Non-linear governing eqaution polar}) are supplemented by six boundary conditions. At the center of the membrane, we require regularity of $d(r,t)$ and $T_{r}(r,t)$
\begin{equation}
\frac{\partial d}{\partial r}=0\quad\mathrm{and}\quad\frac{\partial T_{r}}{\partial r}=0\quad\mathrm{at}\quad r=0.\label{Symmetry BC polar}
\end{equation}
We also assume a zero flux $q$ at the origin, given by $q=r(-h^{3}\partial p/\partial r+\frac{1}{2}hu_{\mathrm{slip}})$,
that under assumptions of finite $u_{\mathrm{slip}}$ and $\partial p_{\mathrm{e}}/\partial r$
at $r=0$ from (\ref{von-Karman stretching polar dim (a)}) implies
\begin{equation}
\lim_{r\to0}rh^{3}\frac{\partial}{\partial r}\left[\frac{1}{r}\frac{\partial}{\partial r}\left(rT_{r}\frac{\partial d}{\partial r}\right)\right]=0.\label{Additional BC for flux}
\end{equation}
At the edge of the membrane, we consider the following boundary conditions 
\refstepcounter{equation}
$$
  d=r\frac{\partial T_{r}}{\partial r}+(1-\nu)(T_{r}-1)=\frac{\partial}{\partial r}\left(rT_{r}\frac{\partial d}{\partial r}\right)=0\quad\mathrm{at}\quad r=R_{\mathrm{m}}.
  \eqno{(\theequation{a,b,c})}\label{Edge of membrane non-dim}
$$
The first two conditions (\ref{Edge of membrane non-dim}\textit{a--b}) correspond to no transverse displacement, and fixed horizontal displacement
at an outer circular frame holding the membrane, respectively. The last condition (\ref{Edge of membrane non-dim}\textit{c})
is obtained, as previously, from the elastic balance (\ref{von-Karman stretching polar non-dim (a)})
by assuming the fluidic and external pressures are both zero at the
boundaries, $p=p_{\mathrm{e}}=0$.

\subsection{Asymptotic analysis for weakly nonlinear effects due to induced tension}
In this section, we use asymptotic analysis to study the weakly nonlinear effects arising from internal tension formed in the elastic sheet during the deflection.
We apply asymptotic expansions to decouple the two-way coupled
nonlinear equations (\ref{Non-linear governing eqaution polar}) and
(\ref{von-Karman stretching polar non-dim (b)}), assuming small deformation
and strong pre-stretching of the elastic membrane, and obtain a correction
for the effect of induced internal tension.

Assuming small elastic deformations, $\beta=\tilde{d}^{*}/\tilde{h}_{0}\ll1$,
we eliminate the hydrodynamic nonlinearity in the Reynolds equation
and obtain 
\begin{equation}
\frac{\partial d}{\partial t}+\nabla_{r}^{2}[\boldsymbol{\nabla}_{r}\cdot(T_{r}\boldsymbol{\nabla}_{r}d)]=-\boldsymbol{\nabla}_{r}\cdot\boldsymbol{f}_{F}+\nabla_{r}^{2}f_{E}.\label{Linearized governing equation deformation polar}
\end{equation}
For the case of strongly pre-stretched elastic membrane, $\alpha=\tilde{T}_{\mathrm{in}}/\tilde{T}\ll1$,
we expand the deformation and the tension in powers of $\alpha$
\begin{equation}
d=d^{(0)}+\alpha d^{(1)}+O(\alpha^{2}),\quad T_{r}=1+\alpha T_{r}^{(1)}+O(\alpha^{2}),\label{Asym d}
\end{equation}
where $\alpha$ is a small parameter satisfying $\mathrm{max}(\epsilon Re,Wo,\epsilon^{2},\beta,\lambda)\ll\alpha\ll1$.
Substituting (\ref{Asym d}) into the coupled governing equations
(\ref{von-Karman stretching polar non-dim (b)}) and (\ref{Linearized governing equation deformation polar}),
results in three one-way coupled linear equations for the
leading-order deformation and the first-order correction
for the deformation and tension fields. In appendix D of the supplementary
material, we present a detailed derivation of the governing equations
and the appropriate boundary conditions at each order, and provide
closed-form solutions for the leading-order deformation
and the first-order correction for the tension.

Similarly to $\mathsection$ 3, as an illustrative example we consider
the non-uniform electro-osmotic slip velocity as a driving force and
specifically focus on a spatially non-uniform actuation
of the form 
\begin{equation}
f_{F}(r,t)=V_{r}J_{1}\left(\frac{\chi_{n}r}{R_{\mathrm{m}}}\right)H(t)\boldsymbol{\hat{r}},\label{Bessel actuation-1-1}
\end{equation}
corresponding to a zeta-potential distribution of $\zeta(r)=rJ_{1}(\chi_{1}r/R_{\mathrm{m}})$,
subjected to a suddenly applied electric field $V_{r}/r$ in the $\boldsymbol{\hat{r}}$
direction, where $J_{1}(\chi_{n}r/R_{\mathrm{m}})$ is the Bessel
function of the first kind and of the first order, and
$\chi_{n}$ is the $n$th root of $J_{0}(\chi_{n})=0$. In
our analysis, we hereafter focus on the first root, $\chi_{1}=2.4048$. 

Using the governing equations and closed-form solutions derived in appendix D of the supplementary material, we can obtain analytical solutions
for $d^{(0)}(r,t)$ and $T_{r}^{(1)}(r,t)$ resulting from (\ref{Bessel actuation-1-1}).
The leading-order deformation field is 
\begin{equation}
d^{(0)}(r,t)=-\frac{V_{r}R_{\mathrm{m}}^{3}}{\chi_{n}^{3}}J_{0}\left(\frac{\chi_{n}r}{R_{\mathrm{m}}}\right)\left[1-\exp\left(-\frac{\chi_{n}^{4}}{R_{\mathrm{m}}^{4}}t\right)\right],\label{LO solution deformation Bessel}
\end{equation}
while the first-order correction for tension distribution reads 
\begin{equation}
T_{r}^{(1)}(r,t)=\frac{V_{r}^{2}R_{\mathrm{m}}^{4}}{4\chi_{n}^{4}}\left[\frac{\nu J_{1}(\chi_{n})^{2}}{1-\nu}+J_{0}(\bar{r})^{2}+J_{1}(\bar{r})^{2}-\frac{J_{0}(\bar{r})J_{1}(\bar{r})}{\bar{r}}\right]\left[1-\mathrm{e}^{-\frac{\chi_{n}^{4}}{R_{\mathrm{m}}^{4}}t}\right]^{2},\label{FO solution tension Bessel-1-1}
\end{equation}
where $\bar{r}=\chi_{n}r/R_{\mathrm{m}}$. The maximum value
$T_{r,{\mathrm{max}}}^{(1)}$ is obtained at the center of the membrane  
\begin{equation}
T_{r,{\mathrm{max}}}^{(1)}=\frac{V_{r}^{2}R_{\mathrm{m}}^{4}}{4\chi_{n}^{4}}\left[\frac{\nu J_{1}(\chi_{n})^{2}}{1-\nu}+\frac{1}{2}\right].\label{Tmax Bessel-1-1}
\end{equation}
While it is difficult to obtain a closed-form solution for transient
first-order deformation $d^{(1)}(r,t)$, the corresponding steady-state
deformation depends solely on the spatial coordinate and admits a
closed-form solution 
\begin{equation}
d_{\mathrm{s}}^{(1)}(r)=\int_{r}^{R_{\mathrm{m}}}T_{r,{\mathrm{s}}}^{(1)}\frac{\mathrm{d}d_{\mathrm{s}}^{(0)}}{\mathrm{d}\xi}\mathrm{d}\xi,\label{FO steady deformation polar-1}
\end{equation}
where the subscript ${\mathrm{s}}$ denotes the steady state. 
\begin{figure}
 \centerline{\includegraphics[scale=1]{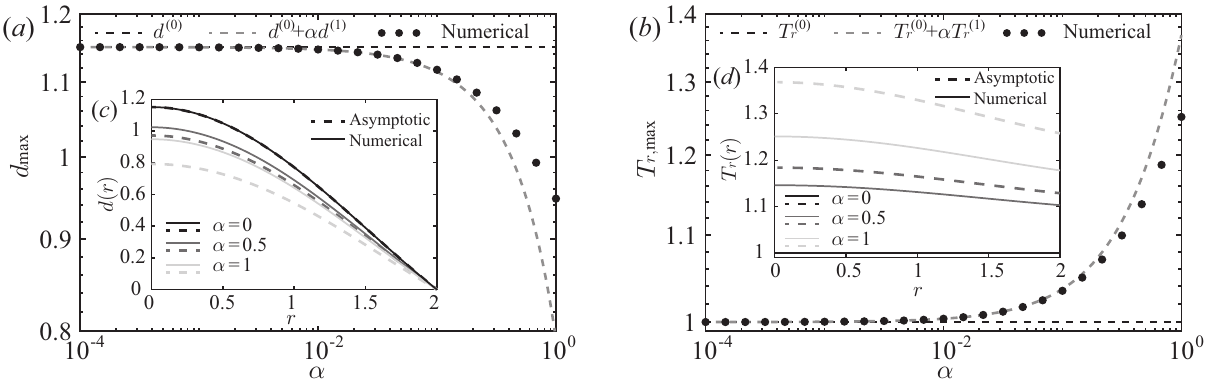}}
\caption{The effect of weak nonlinearity due to induced tension.
(a--b) Comparison between asymptotic and numerical results for the
maximum deformation and tension at steady state as a function of $\alpha$.
Dashed black and gray lines correspond to the leading- and first-order
asymptotic solutions, respectively, while black dots correspond to
the numerical solution. (c--d) Comparison between asymptotic (dashed
lines) and numerical (solid lines) solutions for the steady-state
deformation and tension, respectively, for $\alpha=0, 0.5$ and 1. All calculations
were performed using $\beta=0$, $V_{r}=-2$, $\nu=0.5$ and $R_{\mathrm{m}}=2$.} 
\label{F7}
\end{figure}

Figure \ref{F7} summarizes the effect of weak nonlinearity
due to induced tension on the deformation and tension resulting from
the forcing (\ref{Bessel actuation-1-1}). Figures
\ref{F7}(a--b) present the maximum deformation and tension at steady state
as a function of $\alpha$. Dashed black and gray lines correspond
to the leading- and first-order asymptotic solutions obtained from
(\ref{LO solution deformation Bessel}), (\ref{Tmax Bessel-1-1})
and (\ref{FO steady deformation polar-1}), whereas black dots correspond
to the numerical solution. While the leading-order deformation and
tension are independent of $\alpha$, the first-order correction and
numerical solution clearly show the reduction in deformation magnitude
and the increase in the resulting tension, respectively, as the parameter
$\alpha$ increases. In figures \ref{F7}(c--d) we compare the asymptotic (dashed
lines) and numerical (solid lines) solutions for the steady-state
deformation and tension fields in the case of $\alpha=0, 0.5$ and
1, showing good agreement. As can be inferred from the results of
figure \ref{F7}, for $\alpha\ll1$ the first-order asymptotic solutions (\ref{FO solution tension Bessel-1-1})
and (\ref{FO steady deformation polar-1}) accurately describe the
behavior of the deformation and tension fields, but as $\alpha$ approaches and passes $O(1)$ the asymptotic solutions overpredict them. Nevertheless, $d_{\mathrm{max}}$ and $T_{r,\mathrm{max}}$ continue to decrease/increase monotonically with $\alpha$ and for $\alpha\gg1$  scale like $\alpha^{-1/3}$
and $\alpha^{1/3}$, respectively, as shown by the nonlinear investigation in figure \ref{F9}.

\subsection{Numerical investigation of nonlinear effects}
\begin{figure}
 \centerline{\includegraphics[scale=1]{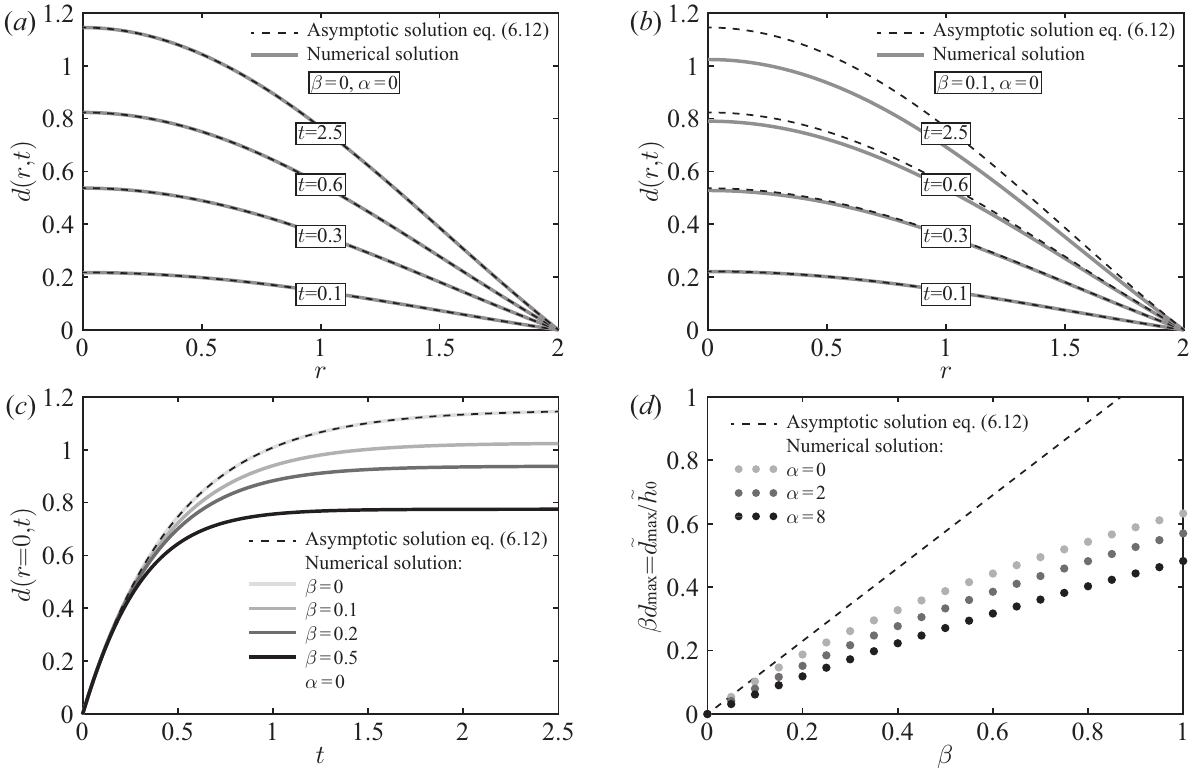}}
\caption{Investigation of nonlinear effects on the transient
behavior and the magnitude of deformation. (a--b) The time evolution
of the deformation field for $\beta=0$ (a) and $\beta=0.1$ (b),
in the case of a strongly pre-stretched elastic membrane, $\alpha=0$.
Gray solid lines represent the numerical results and black dashed
lines represent the leading-order asymptotic solution (\ref{LO solution deformation Bessel}).
(c) The maximum deformation, obtained at the center of the membrane,
as a function of time for various values of $\beta$, with $\alpha=0$.
(d) The maximum deformation, in dimensional form, scaled by the initial
fluid thickness, $\tilde{d}_{\mathrm{max}}/\tilde{h}_{0}=\beta d_{\mathrm{max}}$ as a function of the parameter $\beta$, for various values of $\alpha$.
In (c--d) solid lines and dots represent the numerical results,
respectively, and the dashed black curve represents the asymptotic
solution (\ref{LO solution deformation Bessel}).
All calculations were performed using $V_{r}=-2$, $\nu=0.5$ and
$R_{\mathrm{m}}=2$.}
\label{F8}
\end{figure}
To investigate the nonlinear effects of finite $\alpha$
and $\beta$ on the deformation and tension field, we proceed with
a numerical analysis of the viscous$-$elastic governing equations
(\ref{von-Karman stretching polar non-dim (b)})--(\ref{Non-linear governing eqaution polar}).
To study the effect of hydrodynamic nonlinearity ($h^{3})$ on the
transient behavior and on the magnitude of deformation, 
we consider for simplicity the case of a strongly pre-stretched elastic membrane
with $T_{r}=1$ ($\alpha=0)$ and solve numerically the nonlinear
evolution equation (\ref{Non-linear governing eqaution polar}) for
the deformation using finite differences. In addition, to explore the combined
effect of internal tension and hydrodynamic nonlinearity on the
steady-state behavior, we solve numerically the corresponding steady-state
boundary value problem (\ref{von-Karman stretching polar non-dim (b)})--(\ref{Non-linear governing eqaution polar})
subject to the six boundary conditions (\ref{Symmetry BC polar})--(\ref{Edge of membrane non-dim}) 
using MATLAB's routine \texttt{bvp4c}. Further details of the numerical procedures and their cross validation are discussed in appendix E of the supplementary material.

As shown in figure \ref{F8}(a), we first validate our time-dependent numerical
solver by comparing the numerically determined deformation field
(solid lines) for $\alpha=\beta=0$ with the leading-order asymptotic
solution (\ref{LO solution deformation Bessel}) (dashed lines), showing
very good agreement for all times. Figure \ref{F8}(b) presents the time evolution
of the deformation profile (solid lines), for $\beta=0.1$ and $\alpha=0$.
Figure \ref{F8}(c) shows the time evolution of the maximum deformation, obtained
at the center of the membrane, for $\beta=0,\,0.1,\,0.2,\,0.5$ and
$\alpha=0$. Solid lines represent the numerical results, whereas dashed
lines represent the leading-order asymptotic solution (\ref{LO solution deformation Bessel}). It follows from figures \ref{F8}(b--c) that as $\beta$ increases the resulting maximum deformation in
dimensionless form decreases. This behavior can be explained as follows:
since the viscous resistance scales as $h^{-3}=(1+\beta d)^{-3}$,
the increase in $\beta$ leads to lower internal pressure gradient and thus lower
deformation.

Figure \ref{F8}(d) presents the scaled maximum deformation
$\beta d_{\mathrm{max}}$ as a function of $\beta$, where the dots
represent the numerical results for $\alpha=0,\,2,\,8,$ and the
dashed curve represents the asymptotic solution (\ref{LO solution deformation Bessel}),
corresponding to $\alpha=\beta=0$. We note that when exploring the
nonlinear effects on the resulting magnitude of maximum deformation
it is more convenient to discuss $\beta d_{\mathrm{max}}$ rather
than $d_{\mathrm{max}}$, since the former can be expressed as $\beta d_{\mathrm{max}}=h_{\mathrm{max}}-1=\tilde{d}_{\mathrm{max}}/\tilde{h}_{0}$,
representing the relative magnitude of $\tilde{d}$ in terms of the
initial fluid thickness, $\tilde{h}_{0}$. The numerical analysis
reveals that the scaled maximum deformation $\tilde{d}_{\mathrm{max}}/\tilde{h}_{0}$
increases nonlinearly with $\beta$ throughout the investigated range,
showing a sub-linear behavior, which is more pronounced as $\alpha$
increases. Physically, this means that when $\tilde{d}$ becomes comparable
to $\tilde{h}_{0}$, the dependence of the deformation on the applied electric field is no longer linear due to internal tension and reduction in pressure, resulting in deformation that is lower than predicted by the linear response. However, in the small-deformation and strong pre-stretching
limits, $\tilde{d}_{\mathrm{max}}/\tilde{h}_{0}$ increases linearly
with $\beta$, consistent with the leading-order asymptotic solution
(\ref{LO solution deformation Bessel}) (dashed line). 
Furthermore, figure \ref{F8}(d) shows that, as expected, (\ref{LO solution deformation Bessel}) reasonably predicts the maximum deformation, yielding modest relative
error ($|(d_{\mathrm{max}}^{\mathrm{num}}-d_{\mathrm{max}}^{(0)})/d_{\mathrm{max}}^{\mathrm{num}}|\lesssim12\,\%$)
for up to $\beta=0.1$.  
\begin{figure}
 \centerline{\includegraphics[scale=1]{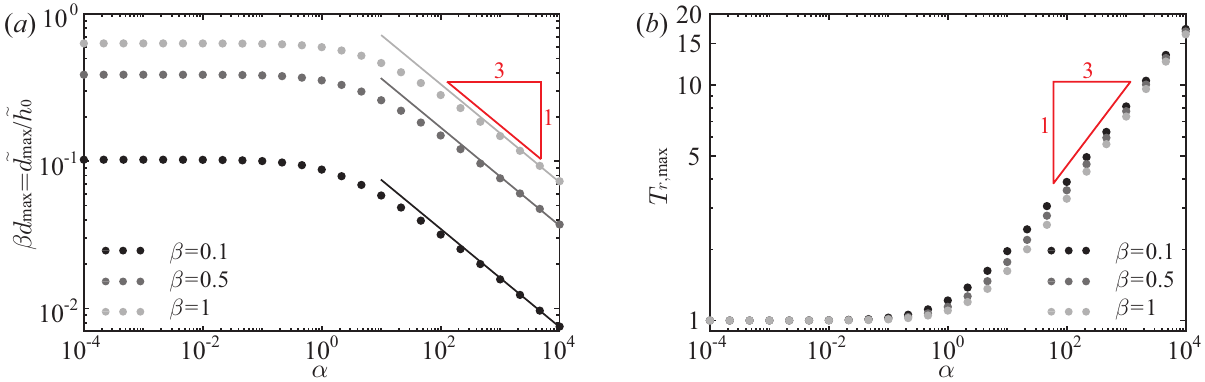}}
\caption{Investigation of nonlinear effect of the induced
tension on the maximum displacement and tension. (a--b) Maximum
displacement, $\beta d_{\mathrm{max}}=h_{\mathrm{max}}-1=\tilde{d}_{\mathrm{max}}/\tilde{h}_{0}$
and tension, $T_{r,\mathrm{max}}$, as a function of $\alpha$, for
$\beta=0.1,\,0.5,$ and 1, obtained from numerical solution of (\ref{von-Karman stretching polar non-dim (b)})--(\ref{Non-linear governing eqaution polar}).
For $\alpha\ll1$, both deformation and tension are independent of
$\alpha$, while for $\alpha\gg1$ the deformation decreases as $\alpha^{-1/3}$
and the tension increases as $\alpha^{1/3}$. All calculations were
performed using $V_{r}=-2$, $\nu=0.5$ and $R_{\mathrm{m}}=2$.}
\label{F9}
\end{figure}

Figure \ref{F9} illustrates the nonlinear effect of induced tension on the scaled maximum deformation and tension for different values of $\beta$. It is evident that at small values of $\alpha$, corresponding to strong pre-stretching, the resulting maximum deformation
and tension are almost independent of $\alpha$, up to $\alpha=O(1)$
(see also figure \ref{F7}). For $\alpha\gg1$ and $\beta\lesssim1$, performing
scaling analysis of (\ref{von-Karman stretching polar non-dim (b)})--(\ref{Non-linear governing eqaution polar})
(with $r\sim1$), we obtain $T_{r}\sim\alpha d^{2}$ and $T_{r}d\sim V_{r}$
thus yielding 
\begin{equation}
d\sim V_{r}^{1/3}\alpha^{-1/3}\quad\mathrm{and}\quad T_{r}\sim V_{r}^{2/3}\alpha^{1/3}\quad\mathrm{for}\quad\alpha\gg1,\,\beta\lesssim1,\label{Scaling with alpha}
\end{equation}
which is consistent with the numerical results shown in figure \ref{F9}.
As expected, for $\beta\lesssim1$ the resulting maximum tension $T_{r,\mathrm{max}}$
indicates weak dependence on $\beta$ over the entire range of values
of the parameter $\alpha$, as shown in figure \ref{F9}(b).

\section{Concluding remarks}
In this work, we examined the effect of pre-stretching and finite boundaries on the elastohydrodynamics of an elastic sheet lying on top of a thin liquid film. Assuming strong pre-stretching and small deformations of the lubricated elastic sheet, we used the linearized Reynolds and F$\mathrm{\ddot{o}}$ppl\textendash von K$\mathrm{\acute{a}rm\acute{a}n}$ equations to derive general analytical solutions describing the deformation in a finite domain due to external forces. These closed-form solutions for realistic configurations  allow one to study the relationship between the magnitude, timescale, and resolution of elastic deformations, as well as to examine spatially discretized actuation. The asymptotic analysis of weakly nonlinear effect due to induced tension for the case of an axisymmetric configuration showed that the first-order asymptotic solutions for the deformation and tension field accurately capture the nonlinear trend even for $\alpha$ of $O(1)$.

We obtained that in the small-deformation ($\tilde{d}\ll\tilde{h}_{0}$)
and strong pre-stretching ($\tilde{T}_{\mathrm{in}}\ll\tilde{T})$
limits, the scaling $\tilde{d}\sim\tilde{\varepsilon}\tilde{\zeta}\tilde{E}\tilde{l}^{*3}/\tilde{T}\tilde{h}_{0}^{2}$
(obtained from (\ref{Slip}) combined with (\ref{N elastic balance}))
is appropriate, showing a linear relation between $\tilde{d}$ and
$\tilde{E}$, and may be used to estimate the resulting deformation.
However, this linear dependence of the deformation on the applied
forcing breaks down as $\tilde{d}$ becomes comparable to $\tilde{h}_{0}$,
indicating a sub-linear behavior, which is more pronounced as $\alpha=\tilde{T}_{\mathrm{in}}/\tilde{T}$
increases.

While our main focus was on actuations applied by the fluid (specifically by non-uniform electro-osmotic flow), the governing equation (\ref{Evoluation calculated-1}) can be readily utilized to investigate the viscous\textendash elastic dynamics due to forces applied directly on the elastic sheet. For such actuation, as shown by the Reynolds equation (\ref{Evoluation calculated}), the viscous flow arises only from temporal variation of the solid deformation field.
Fluid velocity and gauge pressure vanish as the deformation reaches steady state. This is in contrast to the steady state of forcing applied to the fluid, which involves both non-zero fluid velocity and gauge pressure. Furthermore, with appropriate modification of the boundary conditions (e.g. prescribing an external pressure drop), the theoretical model we presented can be readily extended to the study of fluid\textendash structure interaction in elastic microfluidic chips, where micro-channel flow may be driven, for example, by external pressure gradients or peristaltic actuation. 

Actuation at the microscale is currently implemented mostly using MEMS-based technologies, characterized by discrete and rigid elements. The presented results lay the theoretical foundation for implementation of actuation mechanisms based on low-Reynolds-number fluid\textendash structure interaction. While our analysis focused on small deformations, we believe it may be directly useful for the design of soft and continuous actuators. For example, relevant deformations in configurable optics would be on the order of a wavelength (i.e. $<$ 1 $\mu$m in the visible spectrum), and thus could be well described by our model when implemented on a 10 $\mu$m liquid layer. Similarly, the field of microfluidics would highly benefit from configurable microstructures on the order of 10 $\mu$m, which may be implemented on a 100 $\mu$m thick liquid layer.

\section*{Acknowledgments}
This project has received funding from the European Research Council (ERC) under the European Union's Horizon 2020 Research and Innovation Programme, grant agreement no. 678734 (MetamorphChip). We gratefully acknowledge support by the Israel Science Foundation (grant no. 818/13). E.B. is supported by the Adams Fellowship Program of the Israel Academy of Sciences and Humanities. 

\section*{Author contributions}
E.B. performed the theoretical and numerical research. E.B., A.D.G., and M.B. conceived the project, analyzed the results, and wrote the manuscript. R.E., E.B., K.G., and M.B. designed the experiments. R.E. and K.G. fabricated the experimental device. R.E. performed the experiments, analyzed experimental data, and wrote the experimental section.

\bibliographystyle{jfm}
\bibliography{EM}

\end{document}